\documentclass{article}

\usepackage{arxiv}

\usepackage{amssymb}
\usepackage{hyperref}
\usepackage{makecell}
\usepackage{adjustbox,breakurl}
\usepackage{subcaption}

\title{Direct deduction of chemical class from NMR spectra}

\author{Stefan Kuhn\\
Institute of Computer Science, University of Tartu\\
Tartu,Estonia\\
School of Computer Science and Informatics\\
De Montfort University\\
Leicester,United Kingdom\\
\texttt{stefan.kuhn@ut.ee}
\And
Carlos Cobas\\
Mestrelab Research, S.L.\\
Santiago de Compostela,Spain\\
\And
Agustin Barba\\
Mestrelab Research, S.L.\\
Santiago de Compostela,Spain\\
\And
Simon Colreavy-Donnelly\\
School of Computer Science and Information Systems\\
University of Limerick,Ireland
\And
Fabio Caraffini\\
Department of Computer Science\\
Swansea University\\
Swansea,United Kingdom
\And
Ricardo Moreira Borges\\
Instituto de Pesquisas de Produtos Naturais Walter Mors\\
Universidade Federal do Rio de Janeiro,Brazil}

\begin{document}
\maketitle

\begin{abstract}
  This paper presents a proof-of-concept method for classifying chemical compounds directly from NMR data without doing structure elucidation. This can help to reduce time in finding good structure candidates, as in most cases matching must be done by a human engineer, or at the very least a process for matching must be meaningfully interpreted by one. Therefore, for a long time automation in the area of NMR has been actively sought. The method identified as suitable for the classification is a convolutional neural network (CNN). Other methods, including clustering and image registration, have not been found suitable for the task in a comparative analysis. The result shows that deep learning can offer solutions to automation problems in cheminformatics.  
\end{abstract}



\section{Introduction}

Nuclear Magnetic Resonance (NMR) spectroscopy is an established technique in analytical chemistry. As a result of its rich structural and dynamic information content, it is particularly suitable for compound identification. However, a full elucidation may not always be possible or even necessary since some properties might be achievable directly from the spectra. If this is the case, a prioritisation of substances to be closely investigated for compound assignment can be done in the early stages of a study.

A previous example for this idea was demonstrated in \cite{substructures}, where the authors showed that the existence of certain substructures can be concluded from profiles in the spectra. Another potentially useful application is chemical classification. These rely strongly on annotated chemical entities to provide a computable chemical taxonomy based on substructures. In this paper, we infer that chemical taxonomy from the spectra. Similar approach have been applied with Mass Spectrometry (MS) data where chemical classifications can be achieved from fragmentation patterns on MS/MS data \cite{van2016topic}. 

Together with the MolNetEnhancer method \cite{metabo9070144} and Molecular Networks \cite{Nothias2020}, chemical classification is proven to be a valuable tool for compound annotation of unknown compounds and for classification of samples for prioritization for example. The authors have decided to work with spectral images in this current study, but working with raw data would be another option. By working with the images, this frames the problem very much as a digital image processing and machine vision methodology. Computer science has developed a vast range of techniques here, in particular over the last few years, with image processing gaining importance in areas like autonomous driving or security monitoring (see \cite{gonzalez2018digital} for an overview). Whilst these techniques are mostly used to process video or photographic images, they can also be used to process data from scientific instruments \cite{pmid33820092}. Considering this, in this paper we want to examine the question if chemical class can be deducted directly from NMR spectra, using digital image processing techniques. If so, the best method will be identified.

\section{Results}

\subsection{Image similarities}

Table~\ref{table:similarities} shows the results of the image similarity methods described in Section~\ref{sec:convmethods}. First, we can see that, as expected, all methods find the spectra to be identical to themselves. Calculating the similarities within the superclasses and between superclasses gives, on average, similar results. For example, using MobileNetV2 CS gives an average similarity of 0.94 within the superclasses and of 0.91 between superclasses. Although the similarity is slightly lower between superclasses, the difference is not significant. The same is true for the other methods, where SSICompare even gives the same average similarity. From this we conclude that these methods are not suitable to classify the NMR spectra according to the chemical class of the compound.


\begin{center}
\begin{table}
\begin{adjustbox}{max width = \textwidth}
\begin{tabular}{|l|l|l|l|l|l|} 
 \hline
                     &      & \makecell{MobileNetV2 CS} & MobileNetV2 E & ORBCompare  & SSICompare \\ [0.5ex] 
 \hline\hline
With & 1-1 & 1 & 0 & 1 & 1 \\
itself& 5-5  & 1 & 0 & 1 & 1 \\
& 9-9  & 1 & 0 & 1 & 1 \\
\hline\hline
Within    & 1-2  & 0.96            & 8.84          & 0.97   & 0.97  \\
superclass                     & 1-3  & 0.96                & 8.19              & 0.99          & 0.98           \\
                     & 1-4  & 0.96                & 9.21              & 0.99      & 0.98              \\
                     & 5-6  & 0.92                & 12.83             & 0.95           & 0.96             \\
                     & 5-7  & 0.93                & 12.58             & 0.55            & 0.94               \\
                     & 5-8  & 0.94                & 11.30             & 0.75            & 0.96               \\
                     & 9-10 & 0.98                & 6.69              & 0.99          & 0.99              \\
                     & 9-11 & 0.91                & 13.87             & 1.00           & 0.99              \\
                     & 9-12 & 0.96                & 9.63              & 0.90          & 0.98             \\
\hline                     Average & & 0.94 & 10.34 & 0.88 & 0.97  \\
                     \hline\hline
Between  & 1-5  & 0.91            & 13.64       & 0.98   & 0.96     \\
superclasses                    & 1-6  & 0.92                     & 13.27            & 0.93 &  0.98         \\
                     & 1-7  & 0.89      & 15.33& 0.63 &  0.95              \\
                     & 1-8  & 0.93      & 11.96& 0.75 &     0.97         \\
                     & 2-5  & 0.92      & 13.22& 0.97 & 0.97             \\
                     & 2-6  & 0.91      & 13.94& 0.95 &            0.98        \\
                     & 2-7  & 0.90      & 14.92& 0.60 &  0.95       \\
                     & 2-8  & 0.94      & 11.65& 0.82 &          0.97              \\
                     & 3-5  & 0.92      & 13.09& 0.98 &          0.97              \\
                     & 3-6  & 0.94      & 10.89& 0.92 &          0.98              \\
                     & 3-7  & 0.90      & 14.51& 0.66 &            0.96           \\
                     & 3-8  & 0.94      & 11.24& 0.76 &          0.97              \\
                     & 4-5  & 0.91      & 14.18& 0.98 &      0.97             \\
                     & 4-6  & 0.90      & 14.51& 0.94 &          0.98              \\
                     & 4-7  & 0.89      & 15.38& 0.62 &  0.96       \\
                     & 4-8  & 0.93      & 12.26& 0.70 &          0.97 \\
                     & 1-9  & 0.93      & 12.05& 1.00 &          0.98             \\
                     & 1-10 & 0.93      & 12.00& 0.98 &          0.98              \\
                     & 1-11 & 0.87      & 16.50& 1.00 &          0.99              \\
& ... &... &...&...&...\\
\hline
Average & & 0.91        & 12.84   & 0.86   & 0.97 \\
\hline\hline
\end{tabular}
\end{adjustbox}
\caption{Distance measures achieved using various clustering methods. The numbers in the ``Images'' column indicate the images used as explained in Section~\ref{sec:convmethods}. The full table is found in the supplemental materials S1.}
\label{table:similarities}
\end{table}
\end{center}

\subsection{Image registration}

In order to test the image registration, we have trained the CNN model used in VoxelMorph for 20 epochs to ensure convergence with the training images of one class. We then register test images from that class and some other classes. We then sum up pixel by pixel the absolute value of the shifts found by the registration. Similar images should need less change and have therefore a smaller shift overall. An example run is available in the GitHub repository in \textit{results/voxelmorph.txt}. The class used for training here are the Benzoids. The convergence in the training is visible. We then sum up the shifts for three benzoids, the resulting values are 43499, 38834, and 41487. Doing the same for three Organic oxygen compounds gives 50596, 44232, and 44537. For other classes, the results were similar, in some cases the other molecules needed even less change than those from the same class. From those values, we can conclude that no useful information was found by this method. 


\subsection{Clustering based on deep learning}

The authors have executed the clustering as described in Section~\ref{sec:clustermethods} on the 400 images in the training set. Those are from nine superclasses, so we did the clustering with nine clusters. Those nine clusters formed are shown in Table~\ref{table:clusters}, together with the superclasses the compounds clustered in them came from. From this table, we can conclude that the clustering is not conclusive, clusters contain compounds from a variety of classes and the classes distribute over all clusters.

In order to test this, we make an assignment of classes to clusters which optimises the number of correct guesses. The pairs of clusters/classes are Alkaloids and derivatives/9, Benzenoids/8, Lipids and lipid-like molecules/4, Nucleosides, nucleotides, and analogues/7, Organic acids and derivatives/6, Organic nitrogen compounds/6, Organic oxygen compounds/3, Organoheterocyclic compounds/5, Phenylpropanoids and polyketides/2. The sum of correctly classified samples in this combination is 87, which is 21.75\%. This is not much above the value to expect from a random guess. This shows that the clustering does not produce a meaningful result.

\begin{center}
\begin{table}
\begin{adjustbox}{max width = \textwidth}
\begin{tabular}{|c | c c c c c c c c c| c|} 
 \hline
 Actual superclasses/Cluster number & 1 & 2 & 3 & 4 & 5 & 6 & 7 & 8 & 9 & Sum\\ [0.5ex] 
 \hline\hline
 Alkaloids and derivatives & 0 & 5 & 0 & 0 & 2	& 1 &	1 &	0 &	9 & 18 \\ 
 \hline
 Benzenoids & 1 & 17 & 2 & 6 & 11 & 4 & 2 & 4 & 8 & 55\\
 \hline
 Lipids and lipid-like molecules & 0 & 4 & 4 & 23 & 13 & 15 & 3 & 3 & 16 & 81\\
 \hline
 Nucleosides, nucleotides, and analogues & 0 & 1 & 1 & 0 & 0 & 1 & 5 & 0 & 5 & 13\\
 \hline
 Organic acids and derivatives & 6 & 2 & 17 & 13 & 2 & 16 & 6 & 1 & 3 & 66\\  \hline
 Organic nitrogen compounds & 1 & 2 & 0 & 0 & 0 & 8 & 1 & 0 & 0 & 12\\  \hline
 Organic oxygen compounds & 1 & 8 & 14 & 3 & 2 & 24 & 5 & 1 & 4 & 62\\  \hline
 Organoheterocyclic compounds & 0 & 11 & 10 & 4 & 7 & 6 & 6 & 0 & 5 & 49 \\  \hline
 Phenylpropanoids and polyketides & 1 & 11 & 0 & 4 & 3 & 2 & 7 & 1 & 15 & 44\\  \hline
 \hline
 Sum & 10 & 61 & 48 & 53 & 40 & 77 & 36 & 10 & 65 & 400\\  \hline
\end{tabular}
\end{adjustbox}
\caption{The 9 clusters formed by VGG16-based clustering and the actual classes of the members.}
\label{table:clusters}
\end{table}
\end{center}


\subsection{CNN}
\label{sec:cnnresults}

The accuracy achieved using the simple convolutional neural network described in Section~\ref{sec:cnnmethods} is given in Table~\ref{table:cnnresults}. The outputs of the runs are provided in the \textit{results} directory of the GitHub repository.

\begin{table}
\centering
\begin{tabular}{|c | c c c |} 
 \hline
 Accuracy & \makecell{HMBC} & \makecell{HSQC} & \makecell{HMBC and HSQC} \\ [0.5ex] 
 \hline\hline
 Average & 63.23\% & 42.57\% & 44.47\%\\
 Min/Max & 60.00/67.61\% & 38.09\%/46.66\% & 40.00\%/63.80\%\\ 
 \hline
\end{tabular}
\caption{Classification accuracy achieved by the network described in Section~\ref{sec:cnnmethods}, over ten runs.}
\label{table:cnnresults}
\end{table}

The numbers show as the main finding that a CNN is able to distinguish chemical classes by using visual NMR spectra. Since we have 9 classes, a random selection of classes should give about 11\% accuracy. All the accuracies from this method are well above that. HMBC alone has better results than HSQC alone, and both together are similar to HSQC. For single networks, this is in line with the results reported in \cite{substructures} for substructure classification. It should be noted that, as explained in Section~\ref{sec:cnnmethods}, the results could most likely be optimised, and this represents a proof of concept. In contrast to \cite{substructures}, the combined results are not similar to the HMBC results but to the HSQC result.

\section{Methods}

\subsection{The data}

The source of NMR spectra for this article is nmrshiftdb2 \cite{nmrshiftdb2} and the Biological Magnetic Resonance Bank (BMRB) \cite{10.1093/nar/gkm957}. For BMRB, raw NMR data for several small molecules are available for download at \url{https://bmrb.io/ftp/pub/bmrb/metabolomics/entry_directories/}. We have downloaded, using a script, HMBC and HSQC spectra for all compounds for which both spectra exist. The structures were also downloaded in the \texttt{mol} file format. Similarly, nmrshiftdb2 offers a raw data download for some compounds. We have downloaded HMBC and HSQC spectra together with the structures in mol file format here as well, where available. 

All structures were then submitted to the ClassyFire interface at \url{http://classyfire.wishartlab.com/queries/new}. ClassyFire \cite{DjoumbouFeunang2016} is a software that classifies chemical compounds according to a well-defined ontology by structural features. The ontology term used in this article is ClassyFire \textit{superclass}. The use of superclass was mandated because there is a reasonable number of examples of superclasses in the data, whereas more specific terms do not have enough examples for training. We have only used superclasses which have a minimum number of examples of 15. We also performed a random split into training and test data. The final classes and their examples are shown in Table~\ref{table:classes}.

All 2D spectra were processed automatically starting from the raw time domain data using a custom Mnova script. In all cases, one level of zero-filling was applied along the direct dimension (F2). For the indirect dimension (F1), zero-filling was applied so that the number of points is the same as in F2. Linear Prediction was not used. For non-phase sensitive experiments (e.g. COSY), magnitude was calculated and the apodization functions applied were sine bell for F2 and sine square for F1. Phase sensitive experiments (e.g. HSQC), were automatically phase corrected and the apodization functions were sine bell (90º) for both dimensions. As a result, the spectral images have the same scale, no additional rulers, grids, or similar in the image, and the same depth display of the z-dimension. This is different from the approach in \cite{substructures}, where images uploaded to BMRB where used. Those images are not uniform with respect to the settings mentioned. The images used, in directories for the superclasses and training/test set, are available in the \textit{classesbothfinal} folder in the GitHub repository of the project. Such a large set of uniformly processed NMR data, to the authors' knowledge, was not available so far.

\begin{center}
\begin{table}
\begin{adjustbox}{max width = \textwidth}
\begin{tabular}{|c | c c c |} 
 \hline
 Superclass & \makecell{Training\\instances} & \makecell{Test\\instances} & \makecell{Total\\instances} \\ [0.5ex] 
 \hline\hline
 Alkaloids and derivatives & 13/5 & 3/2 & 16/7 \\ 
 \hline
 Benzenoids & 48/7 & 12/0 & 60/7 \\
 \hline
 Lipids and lipid-like molecules & 68/13 & 16/5 & 84/18 \\
 \hline
 Nucleosides, nucleotides, and analogues & 13/0 & 4/0 & 17/0 \\
 \hline
 Organic acids and derivatives & 64/2 & 19/0 & 83/2\\  \hline
 Organic nitrogen compounds & 12/0 & 3/0 & 15/0 \\  \hline
 Organic oxygen compounds & 52/10 & 15/3 & 67/13 \\  \hline
 Organoheterocyclic compounds & 45/4 & 11/2 & 56/6 \\  \hline
 Phenylpropanoids and polyketides & 39/5 & 9/1 & 48/6 \\  \hline
 \hline
 Sum & 354/46 & 92/13 & 446/59 \\  \hline
\end{tabular}
\end{adjustbox}
\caption{Overview of the number of samples for which HMBC and HSQC spectra were available for this project. In the pairs of numbers, the first number refers to BMRB, the second to nmrshiftdb2.}
\label{table:classes}
\end{table}
\end{center}

\subsection{Image similarities}
\label{sec:convmethods}

As a working hypothesis, the authors assume that spectra of compounds of the same class are more ``similar'' than spectra of compounds of different classes. On the basis of this, it should be possible to determine the distance of a spectrum to spectra of compounds of the various ClassyFire superclasses. The most similar spectrum should then tell which class the new compound belongs to. To test this hypothesis, the authors have tested a number of commonly used image similarity metrics.

\begin{itemize}
    \item MobileNetV2 pre-trained ML model by google \cite{https://doi.org/10.48550/arxiv.1801.04381}, combined with cosine similarity (MobileNetV2 CS in here), using the implementation from Keras.
    \item MobileNetV2 pre-trained ML model by google, combined with Euclidean distance (MobileNetV2 E in here), using the implementation from Keras.
    \item ORB key features matching\cite{6126544}, as implemented in OpenCV (\url{https://docs.opencv.org/3.4/d1/d89/tutorial_py_orb.html}).
    \item Structural Similarity Index (SSIM)\cite{1284395}, as implemented in OpenCV (\url{https://docs.opencv.org/4.x/dd/d3d/tutorial_gpu_basics_similarity.html}).
\end{itemize}

These methods were applied using the implementations mentioned. The authors firstly calculated the similarities between two instances of the same images to check the methods, since we expect identity here. Then, we calculated the similarities between HMBC images from the same class and between images from different classes. The average of the similarities in and between classes was calculated. If the methods can do a classification, we would expect the similarities within the classes to be significantly higher than between different classes. The BMRB numbers of the images used and their numbers in Table~\ref{table:similarities} are as follows: Alkaloids and derivatives: bmse001010 (1), bmse001193 (2), bmse001248 (3), bmse001281 (4), Lipids and lipid-like molecules: bmse000317 (5), bmse000394 (6), bmse000478 (7), bmse000484 (8), Organic oxygen compounds: bmse000302 (9), bmse000303 (10), bmse000304 (11), bmse000306 (12). The code for the calculations is available in the \textit{python} folder in the github repository of the project.

\subsection{Image registration}

Image registration in general means transforming different sets of data into one coordinate system. The possible data include images. A typical application would be to align MRI images of different slices of a human brain. For our purposes, we are measuring the amount of transformation needed to align two spectra. Similar to the image similarity calculation, our hypotheses is that spectra for compounds from the same class would require less change to align.

In order to test this we have used the image registration program VoxelMorph \cite{Balakrishnan_2018}. VoxelMorph is based on deep learning and is considered an advanced image registration technique. We train it with the images of one class and then try to register images from the same and other classes to see the changes necessary. The code for the calculations is available in the \textit{python} folder in the GitHub repository of the project.

\subsection{Clustering based on deep learning}
\label{sec:clustermethods}

Clustering techniques can been used to group instances by similarity. Similar to before, if the spectra could be clustered by their similarity, the ClassyFire superclass of a compound could be determined by the cluster its spectra fall into. We use a pretrained network for feature extraction and execute k-means clustering, following \url{https://towardsdatascience.com/how-to-cluster-images-based-on-visual-similarity-cd6e7209fe34}. k-means clustering seems particularly appropriate in this case since the number of clusters needed is known and must be the same as the number of ClassyFire superclasses.


\subsection{CNN}
\label{sec:cnnmethods}

Convolutional neural networks (CNNs) have become a popular choice for many tasks, in particular image processing. Typical tasks here are image classification (e. g. find all images in a set which contain a car) or image segmentation (find all cars in an image). The authors are not covering details of the working and the many applications here, for details and literature references see e. g. the recent review \cite{Alzubaidi2021}. What is important in this context is the ability of CNNs to extract complex information. We have shown in \cite{substructures} that chemical information is accessible for CNNs from image representation of NMR spectra. It has also been shown that CNNs can process raw data of NMR spectra \cite{https://doi.org/10.1002/mrc.5292}. This paper focuses purely on the visual information.

\begin{figure}
\begin{subfigure}{.5\textwidth}
  \centering
  \includegraphics[width=.9\linewidth]{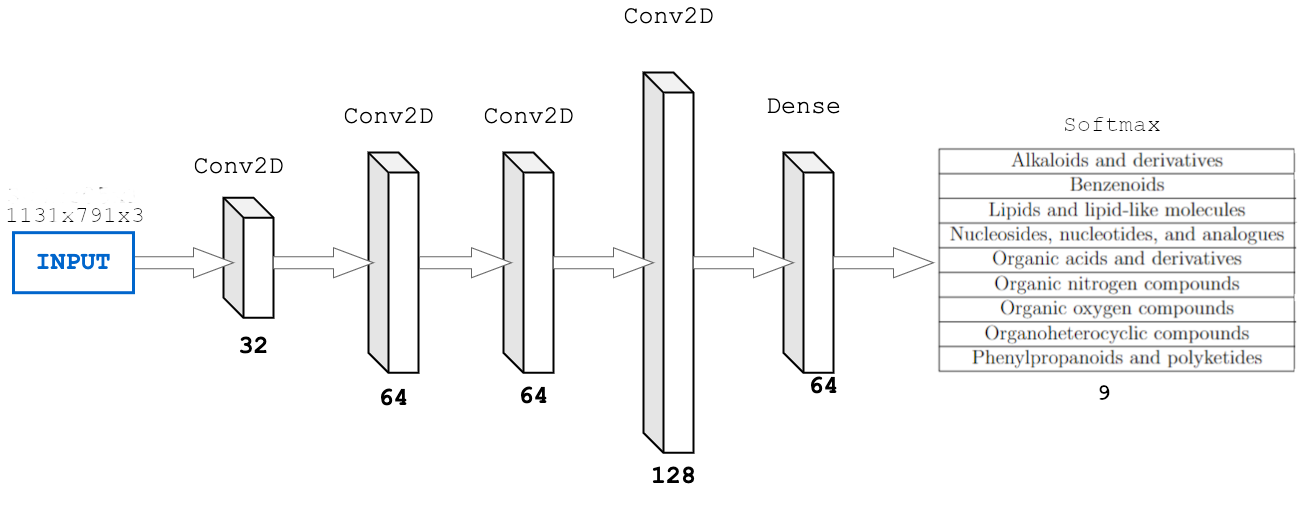}
  \caption{}
  \label{fig:architecture1}
\end{subfigure}%
\begin{subfigure}{.5\textwidth}
  \centering
  \includegraphics[width=.9\linewidth]{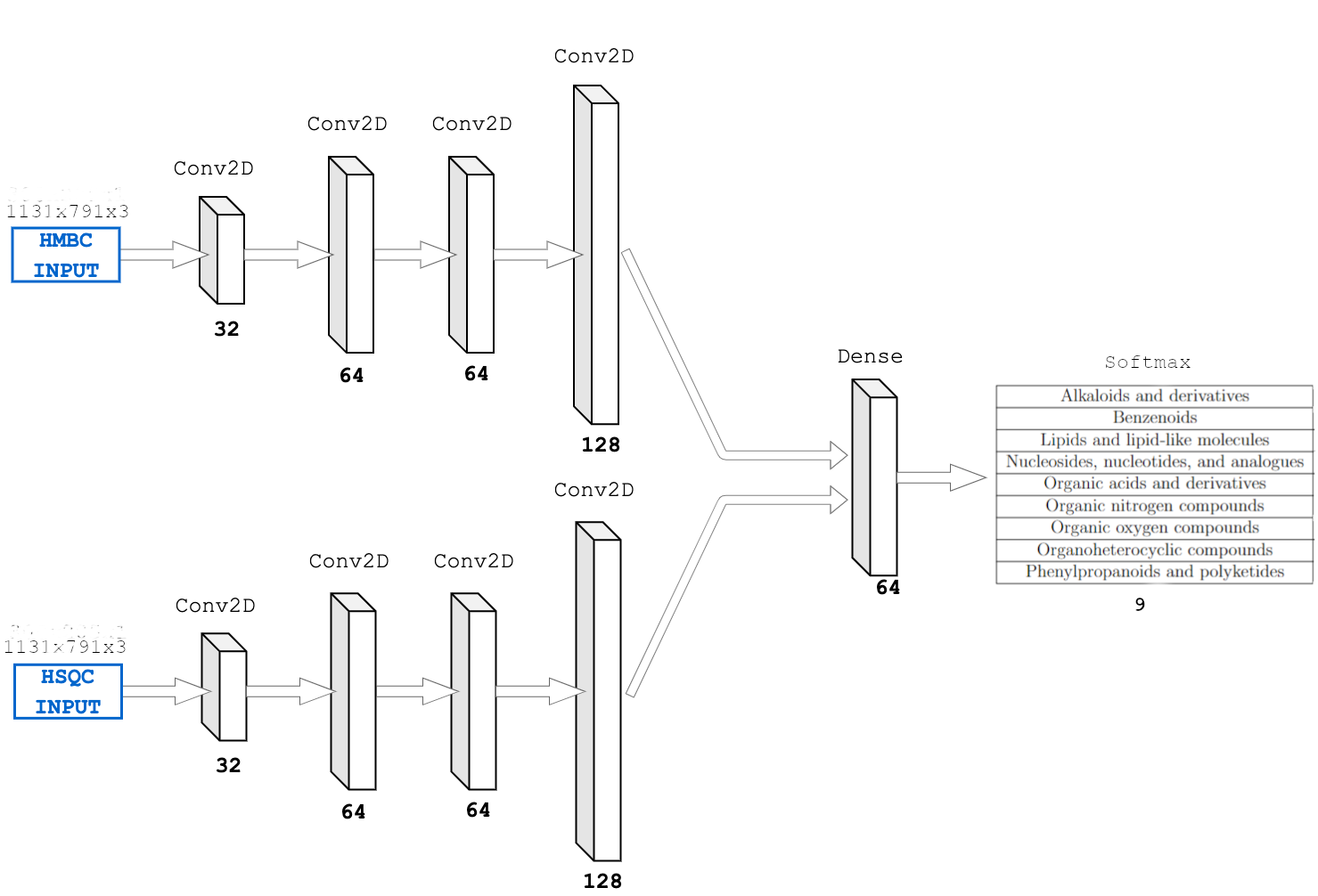}
  \caption{}
  \label{fig:architecture2}
\end{subfigure}
\caption{Architecture of (a) the single spectrum convolutional neural network and (b) the convolutional neural network for combined HMBC and HSQC spectra (from \cite{substructures}).}
\label{fig:architecture}
\end{figure}

The authors used the same networks as in \cite{substructures}. For the HMBC and HSQC only prediction, this is a network consisting of 7 layers, 2 of which are input and output layers (Figure~\ref{fig:architecture1}). This network has been trained with HMBC and HSQC separately. The same structure is also used in the combined neural network (Figure~\ref{fig:architecture2}). This combined CNN architecture consists of 2 initial branches, forming independent CNNs, which are then connected to a dense layer and followed by an output layer. The networks are implemented with Python 3 using Keras and Tensorflow libraries \cite{keras, tensorflow}.

Furthermore, this study assumes a closed-world hypothesis, where all compounds fall in exactly one category (i.e. are of exactly one superclass). The network will decide for one class due to the Softmax activation function in the output layer. Working under an open-world assumption is an open problem in AI research and not part of this pilot study. It should be noted that due to the relatively low number of examples and the this paper being a proof of concept, the authors have not separated a validation set. The authors are using a fixed test set, as opposed to doing cross-validation, which might give slightly different results. The authors also did not optimise the network by tuning its architecture and parameters. Furthermore, we did not try to tune the image export. Images were exported as PNG files with a size of 1133x791 pixels; no further image manipulations were applied. In that sense, the numbers reported in Section~\ref{sec:cnnresults} are a baseline. They represent the average result of ten runs of the training process with the fixed test set. Ten runs were chosen to iron out minor differences between runs. We have trained for 20 epochs for the single networks and for 30 for the combined network, since the networks converge somewhere beyond 10 respectively 20 epochs. The chosen metric is accuracy, since this is most appropriate for a classification problem.

In addition to the CNNs, we have also tried Capsule Neural Networks (CapsNet). These were pioneered by Geoff Hinton et al. in \cite{https://doi.org/10.48550/arxiv.1710.09829} and promise better image processing than CNNs. We have tried the implementations available at \url{https://colab.research.google.com/drive/1WiqyF7dCdnNBIANEY80Pxw\_mVz4fyV-S?usp=sharing} and \url{https://towardsdatascience.com/implementing-capsule-network-in-tensorflow-11e4cca5ecae}. Both are dealing with the MNIST dataset in the original implementation, which is an image of size $28\times28$ pixels. The authors found that dealing with images of our size is impossible for those implementations even on a computer with large memory. Therefore, we have decided not to use CapsNet. This is also justified since a major advantage here is in their ability to deal with rotation of elements, which is not a common distortion in NMR spectra.

\section{Discussion}

The results show that the convolutional neural network is the only method which is able to perform the direct deduction of chemical class from NMR spectra. We can achieve an accuracy of over 60\% with an un-optimized standard CNN. Almost certainly optimizing the network would increase the accuracy even more. In contrast, other methods have not given an accuracy significantly above the random result. This is, perhaps unsurprisingly, true for the conventional image similarities. It is also true for the clustering and image registration methods we have tried. It should be noted that there is a wide range of such methods available and others might give better results, but considering that the CNN is a first attempt as well, this is significant.

In case of CNN, the accuracy is less than that of the same network for substructure classification in \cite{substructures}. This is reasonable since the concept of chemical class as implemented in ClassyFire is more complex than that of a single substructure. As in \cite{substructures}, we get a better result from HMBC spectra than from the HSQC spectra. In this paper, we get an accuracy for the combined spectra that is closer to the HSQC data than the HMBC data. This shows that the method of processing is not suitable here, since the higher information content of HMBC is effectively lost.

\section{Conclusion}

We have shown that CNNs are able to extract chemical class information from NMR spectra. This is in contrast to other image processing methods, which are not able to do so. The accuracies achieved are only intended to show a proof of concept, not as a final result. Improving the network is a task for future research. In particular, the combination of HMBC and HSQC spectra needs attention, since it should get at least as good a result as the best single input result. Another potential research route is the inclusion of more information, e. g. from mass spectrometry.

\section*{Acknowledgement}

SK acknowledges funding by De Montfort University for computational facilities (VC2020 new staff L SL 2020).

\section*{Data statement}

Data and code are available at \url{https://github.com/stefhk3/nmrchemclassify}.


 \bibliographystyle{elsarticle-num} \bibliography{strucclassnmr}

\end{document}